# Space charge formation in chromium compensated GaAs radiation detectors

E. Belas[1*], R. Grill[1], J. Pipek[1], P. Praus[1], J. Bok[1], A. Musiienko[1], P. Moravec[1], O. Tolbanov[2], A. Tyazhev[2], A. Zarubin[2]

[1]Charles University, Faculty of Mathematics and Physics, Institute of Physics, Ke Karlovu 5, Prague 2, CZ-12116 Czech Republic
[2]R&D Center "Advanced Electronic Technologies", Tomsk State University, 36 Lenin av., Tomsk, 634050 Russia

*Corresponding author E-mail: eduard.belas@mff.cuni.cz

**Abstract**

Space charge formation in chromium-compensated GaAs sensors is investigated by the laser-induced transient current technique applying pulsed and DC bias. Formation of non-standard space charge manifested by an appearance of both negatively and positively charged regions in DC biased sensors was revealed during 5 ms after switching bias. Using Monte Carlo simulations of current transients we determined enhanced electron lifetime $\tau = 150$ ns and electron drift mobility $\mu_d = 3650$ cm$^2$/Vs. We developed and successfully applied theoretical model based on fast hole trapping in the system with spatially variable hole conductivity.

**1. Introduction**

Semiinsulating (SI) GaAs compensated by in-diffused chromium (GaAs:Cr) appears nowadays a promising X-ray detector material overcoming known drawbacks of Liquid Encapsulated Czochralski (LEC) SI-GaAs significantly debased by EL2 defect [1-3]. GaAs:Cr exhibits extended electron lifetime and long-term stability at applied bias, conserving simultaneously profitable properties of GaAs-based room-temperature radiation detectors, especially the relatively high average atomic number (Z=32), moderate energy band-gap (Eg=1.42 eV) and high electron drift mobility ($\mu_{de}$=2500 cm$^2$/Vs) [4-5]. One of the key problems for application of SI GaAs in high-performed radiation detectors is still space charge formation resulting in the internal electric field distortion. Model of the space charge formation and electric field distribution in SI LEC GaAs was presented by McGregor [6] and Kubicki [7]. McGregor included the effect of deep levels refilling at high field and Kubicki took the leakage current into account. Model was enhanced by Cola [8], who assumed spectrum of shallow and deep levels for electrons and holes and field enhanced capture cross-section. Experimentally, the electric field distribution in LEC GaAs was investigated using the Transient current technique (TCT) [9-11]. Due to very short lifetime of generated electrons, they measured current transients, which showed anomalous dependence on applied bias significantly warped by effects of associated electronic circuit. Comparison of electric field distribution in LEC GaAs and GaAs:Cr was done by Tyazhev et.al. [12] using Pockels effect. They presented more homogenous distribution of internal electric field in GaAs:Cr compared to LEC GaAs.

In this paper we investigate GaAs:Cr sensors by the laser-induced transient current technique (L-TCT) [13] at pulsed and DC biasing [14]. We focus on the evolution of current waveforms in the time interval 0.1 ms - 5 ms after the switching-on the bias. Observed effects are successfully explained by the model of spatially variable hole conductivity induced by Cr diffusion. Simultaneously, we evaluate the electron lifetime, the electron drift mobility, the charge collection efficiency, evolution of the internal electric field, and the space charge formation. Current waveforms (CWFs) are analyzed by Monte Carlo (MC) simulations following the procedure described in [15]. Specific profile of the electric field, spreading of drifting cloud of electrons by diffusion and surface recombination are newly involved to the simulations. Carrier trapping and detrapping are described by appropriate spatially and temporary constant trapping time $\tau_T$ and detrapping time $\tau_D$. Presented model consistently describes all observed phenomena and its conformity with experimental data is demonstrated. We argue that the internal electric field profile can be easily determined by the L-TCT, which gives direct feedback for the optimization of the growth technology.

## 2. Experimental details

Planar detectors with dimensions 5 mm×5 mm×0.5 mm and one Hall bar sample with dimensions 3 mm×15 mm×0.5 mm were used for detailed characterization of the material by measuring L-TCT and galvanomagnetic properties, respectively. Samples were cut from the SI GaAs:Cr single crystal wafer, which was prepared in the Tomsk State University in Russia by the Cr in-diffusion to both surfaces of the LEC GaAs wafer [1,4]. Samples were mechanically polished using 1 μm $Al_2O_3$ in aqueous suspension and Au/GaAs/Au contacts were prepared by the evaporation. Room temperature bulk resistivity $\rho_0 = 2\times10^9$ Ωcm was deduced from the Hall Effect measurement. The negative sign of the Hall coefficient together with rather low room temperature Hall mobility 1700 $cm^2$/Vs indicate the mixed type of conductivity [16] and an important role of holes in the charge transport. Such findings are in agreement with previously published data [1].

Transient currents were excited through semitransparent detector's cathode by the pulsed laser diode with the above band-gap light at the wavelength 660 nm, which is absorbed in a thin layer ~ 250 nm under the cathode. The laser pulse FWHM was ≈ 0.5 ns. Neutral density filter was used for the laser intensity attenuation to suppress the contribution of photo-generated carriers to the electric field warping and the space charge formation. Photo-generated electrons created near the cathode drift toward the anode and induce transient current whilst holes are immediately collected at the cathode and their effect may be neglected. The signal was amplified and recorded by a digital sampling oscilloscope at the frequency bandwidth 3 GHz [13]. Laser and bias pulses are time-correlated according the scheme outlined in Fig. 1. Pulsing conditions are defined by three parameters; laser pulse delay (LPD), bias pulse width (BPW) and depolarization time (DT). Variable LPD allows to evaluate the progress of the internal electric field profiling through its effect to the current waveform shape. The dynamics of the space charge formation is derived through the Gauss's law.

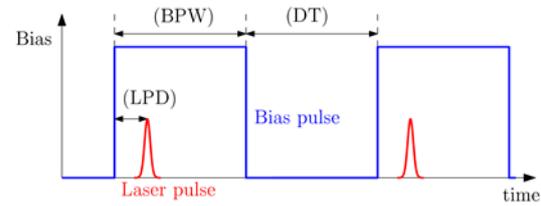

Fig. 1. Scheme of the time-correlation between the pulsed biasing and laser excitation.

## 3. Results and Discussions

Fig. 2 shows bias dependence of electron CWFs measured in planar GaAs:Cr detector using pulse biased L-TCT, where pulsed bias parameters LPD=80 μs, BPW=10 ms and DT=10 ms were chosen. The short LPD assures that there is no enough time for the development of the space charge in the detector within the single bias pulse. Simultaneously, a conceivable residual space charge, which could appear during the pulse, is drained away at the period of DT.

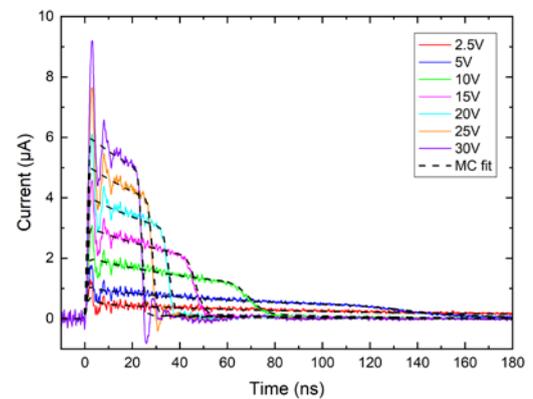

Fig. 2. Bias dependence of electron current waveforms using pulse biased L-TCT. Dashed lines represent the MC fit.

We can thus consider the detector nearly neutral with constant electric field distributed within the whole sample. Consequently, the slope of CWF is affected only by the photo-electron trapping and recombination [14]. The sharp drop of CWFs corresponding to the transit time of electrons drifting toward the anode was determined for the first time in this material. High-frequency oscillations apparent at the initial stage of CWFs are caused by response of the electronic circuit to the fast rising edge of the current waveform and they are not analyzed in this paper. We also observed a weak hole signal using anode illumination in the pulsed bias conditions, but its structure was too fuzzy that we could not determine transit time and other transport characteristics. The Monte Carlo fit plotted by dashed lines in Fig. 2 was obtained assuming the constant electric field across the sample thickness and defect model with two electron traps [15]; one shallow trap characterized by trapping and detrapping time $\tau_{TS}$=250 ns and $\tau_{DS}$=40 ns, respectively, and one deep trap with trapping time $\tau_{TD}$=150 ns and negligible detrapping. Electron drift mobility evaluated from the MC fit of each CWF at pulse bias conditions resulted $\mu_{de}$=3650 cm$^2$/Vs.

An important finding is documented in Fig. 3, where identical CWFs presented in Fig. 2 normalized by applied bias are plotted. Nearly equal course of the Current/Bias curves in its transient period proves the linear scaling of the initial current transient with bias in the pulsed bias. This fact testifies to the bias-independent charge drifting through the sample and in consequence to the negligible electron surface recombination at the cathode.

measured in undoped n-GaAs by the Hall effect measurement [17]. Therefore an existence of a shallow trap with very fast trapping/detrapping on a sub-ns scale can be considered together with enhanced charged impurity scattering.

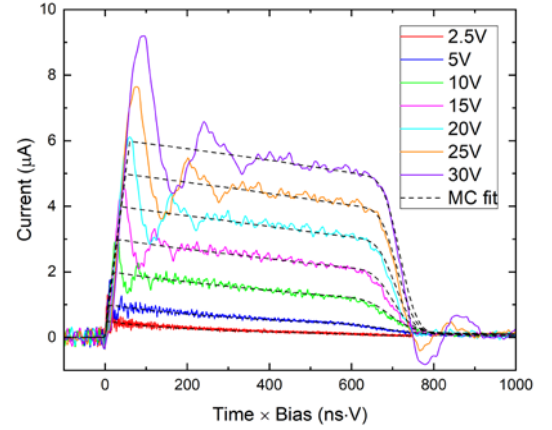

Fig. 4. Bias dependence of electron current waveforms on biased time (Time × Bias) using pulsed bias. Dashed lines represent the MC fit.

The L-TCT CWFs measured at DC bias are presented in Fig. 5.

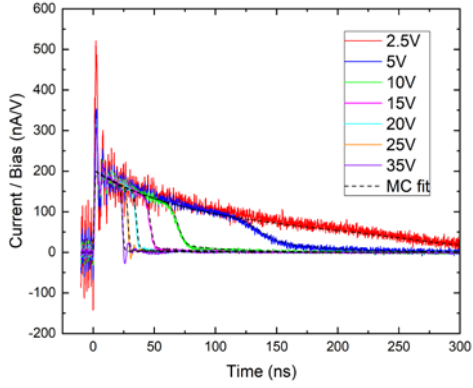

Fig. 3. Bias dependence of normalized electron current waveforms using pulsed bias. Dashed lines represent the MC fit.

Bias dependence of CWF profiles presented in Fig.2 on the biased time, Time × Bias, is shown in Fig. 4 for demonstration of linear scaling of the transit time with applied bias. We see an excellent agreement of the MC fit with waveforms at all biases. The effect of shallow electron trap level is evident mainly in low bias (2.5-10V), where it increases and broadens the tail of CWFs around the transit time. Without shallow level these waveforms would decrease more rapidly and the fit would worsen. In contrast to (CdZn)Te [15] where shallow traps with low $\tau_{TS}$ and $\tau_{DS}$ were reported and carrier drift was delayed by significant trapping, the shallow level defined in this paper does not visibly reduce the electron mobility. The trapping time of shallow level is comparable with the transit time even at the lowest biasing and the most of electrons reach the anode without trapping. Although evaluated electron drift mobility is higher than value presented in [4-5], it is significantly less than the electron mobility $\mu_{He}$=9400 cm2/Vs

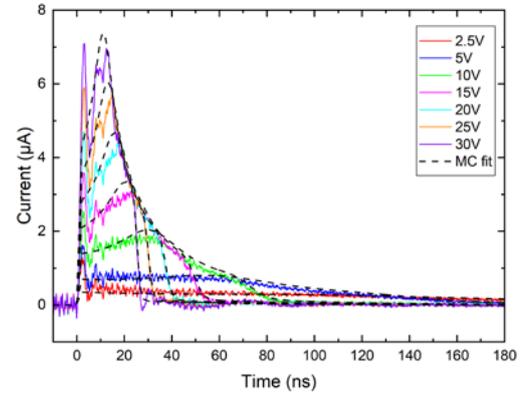

Fig. 5. Bias dependence of electron current waveforms using DC biased L-TCT. Dashed lines represent the MC fit.

We got an unusual shape of CWFs, which we never observed in other materials or found in the literature. We clearly see that regardless of the magnitude of the biasing, CWFs increase initially and after reaching the maximum approximately at half of the electron transit time they decrease until electrons reach the anode. Such observation may be solely interpreted as an electron transport through the sample permeated by an ascending electric field, i.e. formation of the negative space charge region localized in the layer adjacent at the thickness about L/2 to the cathode, and by a descending electric field induced by the positive space charge in the part near the anode.

The bias dependence of collected charge evaluated by the integration of CWFs in both bias conditions presented in Fig. 2 and Fig. 5, is shown together with the Hecht equation fits [18] in Fig. 6. We evaluated the electron mobility-lifetime products $\mu_e\tau_e^{Pulse}$ =5.5×10$^{-4}$ cm$^2$/V and $\mu_e\tau_e^{DC}$ =5.2×10$^{-4}$ cm$^2$/V from the pulse and DC biased CWFs, respectively. The mobility-lifetime product in pulsed regime $\mu_e\tau_e^{Pulse}$ is equal to the product of $\mu_{ed}$ = 3650 cm$^2$/Vs and $\tau_{TD}$ = 150 ns obtained by the MC simulation. Let us note that the shallow electron trap does not induce real losses to the drifting charge and it does not affect the charge collection. Consequently, the electron lifetime is equal to $\tau_{TD}$. With respect to the very low surface recombination, $\mu_e\tau_e^{Pulse}$ represents the right value of $\mu_e\tau_e$ unaffected by the space charge formation. Slightly lower value of $\mu_e\tau_e^{DC}$ is caused by the electric field warping in DC conditions due to the space charge formation in the detector and subsequently by extended transit time. For a comparison, applying pulse height spectrum analysis using alpha particles we obtained $\mu_e\tau_e^{alpha}$ = 1.4×10$^{-5}$ cm$^2$/V. Large difference between $\mu_e\tau_e^{Pulse}$ and $\mu_e\tau_e^{alpha}$ is caused by a strong plasma effect inducing strong loss of photo-generated electrons and the charge collection depression especially at low biases. Incorrect $\mu_e\tau_e^{alpha}$ value was also measured in [12] with the same interpretation.

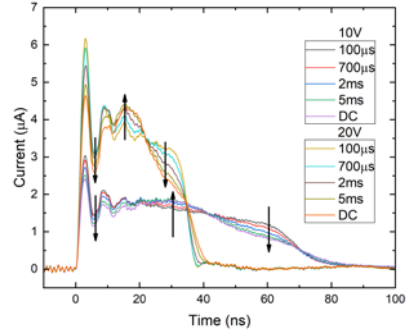

Fig. 7. Electron L-TCT CWF dependence on the laser pulse delay for 10 V and 20 V pulsed biasing. Arrows show the direction of the waveform evolution.

We found that the CWF shapes continuously evolve after the biasing reaching the final form at LPD ≈ 5 ms. We clearly see that CWFs evolve continuously in their whole profile, which proves that both positive and negative space charge regions are formed simultaneously.

The explanation of the parallel formation of both positive and negative space charge regions is a challenging task. Based on our previous investigations of the space charge formation in biased detectors [19] we conclude that obvious models explaining the space charge formation by appropriate contact properties cannot be applied in this material. We need to search for a new model especially comply with the following results and known properties of SI GaAs:Cr material.

1. We measured significantly lower Hall mobility using Hall effect measurement in contrast to the electron drift mobility evaluated with the L-TCT, which testifies on the mixed type of the conductivity of GaAs:Cr and an important effect of holes at the transport. Comparable electron and hole conductivity $\sigma_e \approx \sigma_h$ may be assumed.
2. Very weak L-TCT signal of holes was detected reflecting their very short lifetime and a significant hole trapping. Characteristic lifetime of holes $\tau_h$ < 3 ns was estimated. Recently, $\tau_h$ = 1.4 ns was reported [5] supporting such expectation.
3. SI GaAs:Cr wafer is prepared by Cr in-diffusion from the Cr layer applied to the both surfaces of the wafer. Due to this method, the residual variation of defect structure along the sample thickness can be assumed. Since Cr acts as an acceptor in GaAs:Cr, reduced/enhanced electron/hole density should appear near the surface in contrast to the middle of the detector. In other words, the Fermi energy $E_F$ may be shifted towards the valence band in the subsurface regions relatively to the middle of the detector. Variations at the resistivity similar to those considered here were presented in [1].

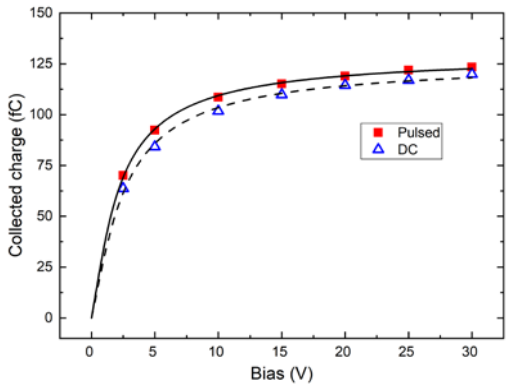

Fig. 6. Bias dependence of collected charge evaluated by integration of CWFs at pulse and DC bias conditions. Solid and dashed lines represent the Hecht equation fits.

The temporal evolution of CWFs measured at selected biases 10 V and 20 V by L-TCT with different LPD (as defined in Fig. 1) is plotted in Fig. 7.

At the search for a theory consistently explaining the above-listed properties we developed a model considering the weak deviation of the hole conductivity along the detector. Since the mechanism of the Cr diffusion and defect compensation is not known in detail in GaAs:Cr yet, we express the course of hole conductivity profiled along the sample thickness by a nearly symmetrical trial function as follows

$$\sigma(x) = \sigma_0 + \sigma_1 \frac{(x-x_m)^2}{\sqrt{(x-x_m)^2+\gamma^2}}, \qquad (1)$$

where $\sigma_0$, $\sigma_1$, $\gamma$, and $x_m$ are model parameters. Anticipating diffusion of Cr occurring from both sides of the wafer, $x_m \approx L/2$ is foreseen. Similar course may be deduced from the profile of resistivity [1]. The cathode is set to $x=0$ and sample thickness $x=L=0.5$ mm.

Let us note that involving holes in the concept is important due to their short lifetime and consequently fast interaction with principal trap levels by the trapping and detrapping. The fast exchange of holes between the valence band and the hole trap assures the local equilibrium between the band and trap, which is critical for the validity of the concept of variable conductivity in this model. Considering the mean free path of holes $< 7$ µm, which appears sufficiently short in the 500 µm thick sample, estimating the hole mobility $\approx 200$ cm$^2$V$^{-1}$s$^{-1}$ and maximum electric field $\approx 1$ kV/cm, we obtain the lifetime of holes $< 3$ ns. Evidently, the shorter lifetime, the better for the validity of the approach. Fast electrons with the extended lifetime could not inspire observed formation of abruptly changing positive and negative space charge.

Equation (1) is combined with obvious Ohm's law

$$j = \sigma E, \qquad (2)$$

Gauss's law

$$\frac{\partial E}{\partial x} = \frac{\varrho}{\varepsilon} \qquad (3)$$

and continuity equation

$$\frac{\partial \varrho}{\partial t} = -\frac{\partial j}{\partial x}, \qquad (4)$$

where $j$, $E$, $\rho$, and $\varepsilon$ are hole current density, electric field, charge density and permittivity, respectively. The principle of the model consists in the charging of the hole trap induced by the variable conductivity and nonequilibrium hole distribution in biased sample. Eq. (1) represents only the hole conductivity, which is thus lower than the total conductivity involving also electrons.

Eqs. (1)-(3) were substituted into eq. (4), which was numerically time-integrated for given $\sigma(x)$. $E(x)$ was calculated by the integration of eq. (3) after each integration step with the boundary condition preserving the bias

$$U = -\int_0^L E(x)dx. \qquad (5)$$

Basic properties of GaAs used at the calculations were taken from [17] and we used the hole mobility $\mu_h=171$ cm$^2$/Vs [3]. Acquired $E(x)$ has entered in the MC simulation of experimental CWFs presented in Fig.5 whilst the trapping/detrapping time of electron traps remained the same as obtained from the fit at pulsed bias shown in Fig. 2. The self-consistent loop of the electric field calculations and MC simulations was repeated converging to the optimum fit. All L-TCT CWFs measured at applied bias ranging in the interval 2.5 V – 30 V, laser pulse delay LPD in the interval 80 µs – 5 ms and illuminating alternatively both sides of the sample were fitted. The unique set of parameters was retrieved from the fitting of electron CWFs as follows: $\sigma_0=2.2\times10^{-10}$ Ω$^{-1}$cm$^{-1}$, $\sigma_1=1.3\times10^{-8}$ Ω$^{-1}$cm$^{-1}$, $\gamma=0.1$ mm and $x_m= 0.225$ mm. Fig. 8 plots the MC fit of L-TCT CWF dependence on the LPD for 20 V pulsed and DC bias. Internal electric field profile in the detector is presented in the inset, where arrows show the direction of the electric field evolution.

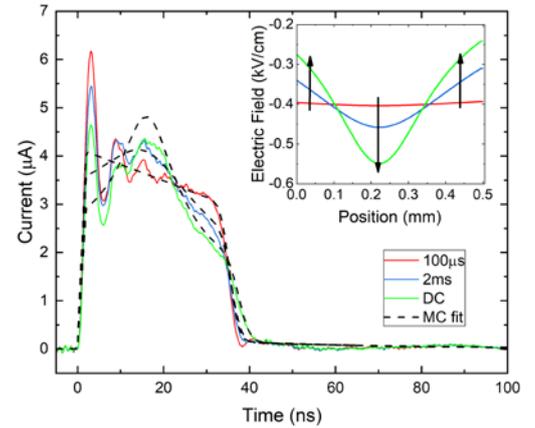

Fig. 8. Electron CWF dependence on the laser pulse delay for 20 V. Dashed lines represent the MC fit. The internal electric field profile in the detector is presented in the inset, where arrows show the direction of the electric field evolution.

The bias dependence of the internal electric field profile representing the theoretical fit of DC CWFs is shown the Fig. 9. We may see an enhanced profiling of the field at increased bias symmetrically modulated by the variable conductivity $\sigma(x)$ shown in Fig.9 as well. Except the sign, the modulation is independent of the polarity of bias, which entails the same character of CWFs measured from both side of the sample.

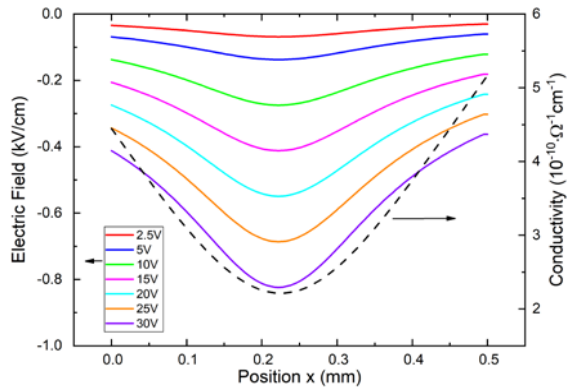

Fig. 9. Bias dependence of the internal electric field profile at DC bias. Dashed line shows the profile of fitted conductivity.

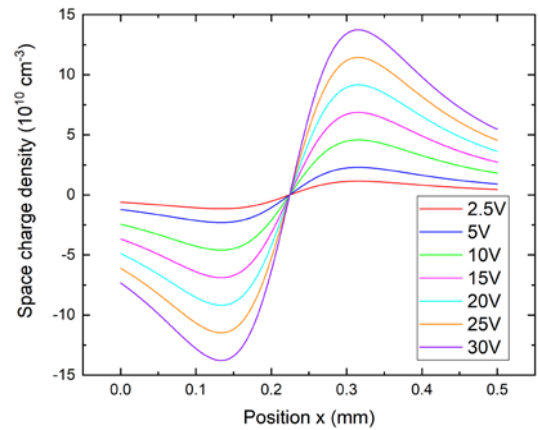

Fig. 10. Bias dependence of the normalized space charge density ($\rho/e$) profile at DC bias.

Considering obtained results, the conductivity profile reaches its minimum at $x_m$= 0.225 mm, where $\sigma(x_m) = \sigma_0 = 2.2\times 10^{-10}$ $\Omega^{-1}$cm$^{-1}$ and maximum at the surface $\sigma(0)= 4.4\times 10^{-10}$ $\Omega^{-1}$cm$^{-1}$; $\sigma(L)= 5.2\times 10^{-10}$ $\Omega^{-1}$cm$^{-1}$. Respective position of Fermi energy $E_F(x_m) = E_V + 0.689$ eV, $E_F(0)= E_V + 0.672$ eV, and $E_F(L)= E_V + 0.668$ eV. The hole and electron densities range from $8\times 10^6$ cm$^{-3}$ to $1.8\times 10^7$ cm$^{-3}$ and from $6\times 10^4$ cm$^{-3}$ to $1.4\times 10^5$ cm$^{-3}$, respectively, which is consistent with negative sign of the Hall coefficient with reduced measured Hall mobility effected by mixed transport.

The steady state space charge density distribution deduced through eq. (3) from the electric field profile presented in Fig.9 is shown in Fig. 10. The principle of the positive charging stems from the weak injection of holes from the anode related with the larger hole conductivity and density in that region. Oppositely, the negative charge appears in the region adjacent to the cathode where the holes are depleted due to relatively lower conductivity in the middle of the sample, which inhibits the hole current. It is worth pointing out to the fact that only two parameters $\sigma_0$ and $\sigma_1$ allowed us to describe both the velocity of the space charge formation and the final magnitude of the space charge.

Simultaneously, the fitted average hole conductivity of the sample $<\sigma> = 3.2\times 10^{-10}$ $\Omega^{-1}$cm$^{-1}$ is slightly lower than the experimentally determined conductivity $5\times 10^{-10}$ $\Omega^{-1}$cm$^{-1}$, which is in agreement with the model prediction above.

## 4. Conclusion

We measured electron current transients in GaAs:Cr sensor using L-TCT in pulsed and DC bias and determined the time evolution of the space charge formation. At the time scale of 5 ms we observed the simultaneous gradual formation of both negative space charge localized near the cathode and the positive space charge near the anode. Electron lifetime $\tau$ = 150 ns and electron drift mobility $\mu_d$ = 3650 cm$^2$/Vs were evaluated from the Monte Carlo simulations. All experimental results have been consistently explained by the model of variable hole conductivity caused by chromium in-diffusion. For the elimination of the internal electric field distortion, pulsed bias application was used. We also showed that using L-TCT the internal electric field profile can be determined giving direct feedback for the optimization of the growth technology. We demonstrated that L-TCT represents the non-destructive method for testing of transport properties in GaAs:Cr as well as in other radiation detector materials with much better precision than it may be reached by other commonly used techniques.

## Acknowledgements

This work was supported by the Grant Agency of the Czech Republic under contract No. 18-12449S and by the Grant of Charles University under contract No. SVV-2019-267306.